# Absence of significant spin current generation in Ti/FeCoB bilayers with strong interfacial spin-orbit coupling

Lijun Zhu[*], Robert A. Buhrman
*Cornell University, Ithaca, New York 14850, USA*
[*]lz442@cornell.edu

After one decade of the intensive theoretical and experimental explorations, whether interfacial spin-orbit coupling (ISOC) at metallic magnetic interfaces can effectively generate a spin current has remained in dispute. Here, utilizing the Ti/FeCoB bilayers that are unique for the negligible bulk spin Hall effect and the strong tunable ISOC, we establish that there is no significant charge-to-spin conversion at magnetic interfaces via spin-orbit filtering effect or Rashba-Edelstein(-like) effect even when the ISOC is stronger than that of a typical Pt/ferromagnet interface and can promote strong perpendicular magnetic anisotropy. We also find a minimal orbital Hall effect in 3*d* Ti.

Efficient generation of spin current is fundamental for the exciting field of spin-orbitronics. It is well known that spin current can arise from the bulk spin Hall effect (SHE)[1,7,8], the topological surface states [9,10] the anomalous Hall effect [11,12], the planar Hall effect [13,14], the spin Seebeck effect [15], spin pumping [16], or non-centrosymmetric crystals [17]. Triggered by the fundamental importance for the condensed matter physics and by the great promise for memory [1-5] and logic technologies [6], exploring novel mechanisms for efficient spin current generation has become a very hot topic. Recently, interfacial spin-orbit coupling (ISOC) of a normal metal/ferromagnet (NM/FM) interface was also proposed to generate non-local spin current and thus dampinglike spin-orbit torque (SOT) [18-24].

In the first mechanism known as spin-orbit filtering effect [18], interfacial spin Hall effect [19,20], or three-dimensional model of ISOC [21], if carriers are scattered across the NM/FM interface by strong ISOC [18-21], the transmitted current could become transversely spin polarized and exert a dampinglike SOT on the adjacent FM layer (Fig. 1). Moreover, when the carriers are scattered by ISOC along the NM/FM interface, it leads to the same symmetry as two-dimensional (2D) Rashba-Edelstein effect [22] and is sometimes known as "Rashba-Edelstein-like" effect, although the Rashba-Edelstein effect is strictly valid only in 2D electron gas systems, not metallic interfaces [25-27]. While theories have predicted that such Rashba-Edelstein-like effect should generate a small fieldlike torque but negligible *charge-to-spin conversion* and dampinglike SOT at metallic *magnetic interfaces* of thin-film systems [25-28], it is experimentally still rather controversial. This is mainly because of the heated experimental debate on the presence of *spin-to-charge conversion* at *non-magnetic* interfaces [29-31](e.g. at Bi/Ag interfaces [32,33]), which is the inverse process of the Rashba-Edelstein-like effect.

In addition, orbital current generated by the orbital Hall effect (OHE) (e.g. in $CuO_x$) can be converted into spin current via spin-orbit coupling of an interface or a FM [23,24]. There is also a report of an "intrinsic Berry curvature"-induced interfacial dampinglike SOT at the $CuO_x$/$Ni_{81}Fe_{19}$ interface [34], however, efficiency of that torque is 2~3 orders of magnitude smaller than that from the bulk SHE of Pt [7] and therefore cannot yield an efficient charge-to-spin conversion.

While these proposals have significantly enriched the scope of the potential mechanisms for spin current generation, clean experimental exam for the effectiveness of these ISOC effects as a function of the ISOC strength is urgently required, which is particularly so when considering the more than one decade of remarkable interests in exploring interfaces as promising spin-charge conversion candidates [18-33]. In most spin Hall metal/FM heterostructures, it is, however, a great experimental challenge to disentangle the amplitude and the polarization orientation of interface-generated spin currents due to the coexistence of the strong spin current generation by the bulk SHE [35-37] and the ISOC-induced loss of spin memory to the lattice [38-40]. It is also a major challenge to obtain a strong and widely tunable ISOC within a same material system without using a strong spin Hall metal.

In this Letter, we quantitatively investigate the potential effects of ISOC in Ti/FeCoB bilayers which are unique because of the minimal bulk SHE [41], the negligible bulk SOC in Ti, and a tunable ISOC at the same time. By varying the Ti thickness (*d*), the strength of the ISOC at Ti/FeCoB interface is tuned from negligibly small to greater than that of typical Pt/FeCoB and Pt/Co interfaces. From this ability, we establish that regardless of *d* and the ISOC strength both dampinglike and fieldlike SOTs in the Ti/FeCoB bilayers are at least two orders of magnitude smaller than that provided by the bulk SHE of Pt, indicating no significant spin current generation by the spin-orbit filtering effect, the Rashba-Edelstein effect, or orbit-spin conversion at Ti/FeCoB interfaces.

For this work, we sputter-deposited perpendicularly magnetized bilayers of Ti 1.7-6.9/FeCoB 1 (the numbers are layer thickness in nm; FeCoB = $Fe_{0.6}Co_{0.2}B_{0.2}$) and in-plane magnetized Ti 1.7-6.9/FeCoB 2 onto oxidized Si substrates. Ti is chosen to provide ISOC at the bottom surface of the FeCoB. Each sample is capped with a 2 nm MgO and a 1.5 nm Ta layer that is oxidized upon exposure to air (see the cross-sectional scanning transmission electron microscopy and electron energy loss spectrum results in [42]). Each layer was deposited at a low rate (e.g. 0.006 nm/s for Ti and FeCoB), with the argon pressure of 2 mTorr. The base pressure is ~$10^{-9}$ Torr. Each sample was patterned into 5×60 μm$^2$ Hall bars for measuring the SOTs and the ISOC by harmonic Hall voltage response (HHVR) measurements [43-45]. Each sample underwent an annealing at 115 °C for 2 minutes during photolithography process. All measurements were performed at room temperature.

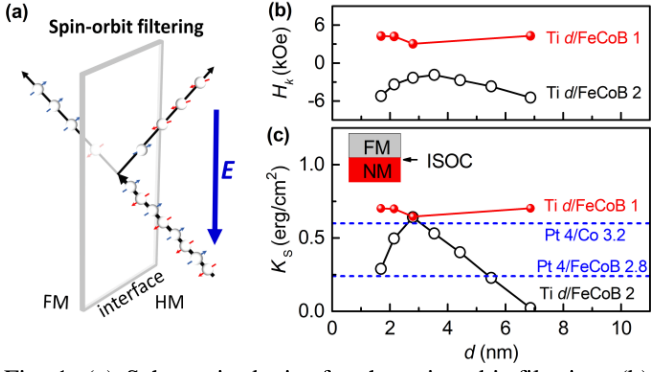

Fig. 1. (a) Schematic depict for the spin-orbit filtering. (b) Perpendicular magnetic anisotropy field ($H_k$) and (c) Interfacial magnetic anisotropy energy density for the Ti/FeCoB interface ($K_s^{ISOC}$) plotted as a function of the Ti thickness. The error bars are smaller than the symbols in (b) and (c). The two dashed blue lines in (c) indicate the $K_s^{ISOC}$ values of in-plane magnetized Pt 4/Co 3.2 [50] and Pt 4/FeCoB 2.8 [39], respectively.

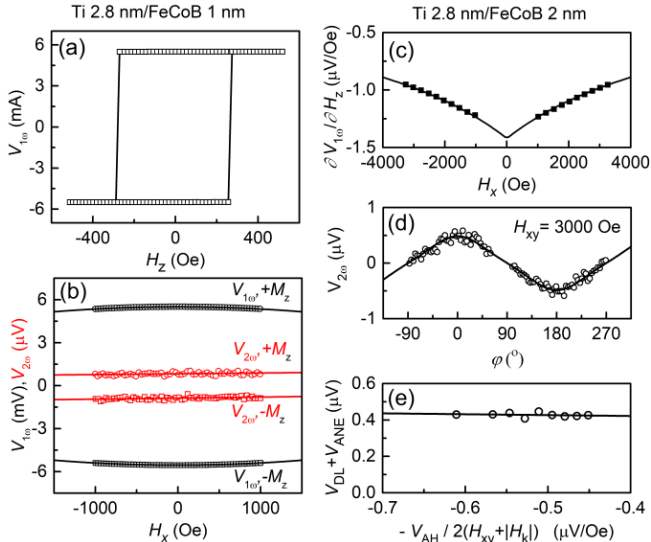

Fig. 2. (a) $V_{1\omega}$ vs $H_z$ and (b) $V_{1\omega}$ and $V_{2\omega}$ vs $H_x$ for the perpendicularly magnetized Ti 2.8/FeCoB 1. (c) $dV_{1\omega}/dH_z$ vs $H_x$, (d) $V_{2\omega}$ vs $\varphi$ ($H_{xy}$ = 3000 Oe), and (e) $V_{DL} + V_{ANE}$ vs $-V_{AH}/2(H_{xy}+|H_k|)$ for the in-plane magnetized Ti 2.8/FeCoB 2. In (b) the red straight lines represent the best linear fits, and the black parabolic lines denote the fits of the data to Eq. (1). In (c), the black curve is the best fit of data to Eq. (2). In (d) the lines represent the best fits of data to Eq. (5). The straight line in (e) represents the best linear fit.

According to Bruno's model [46,47], the interfacial perpendicular magnetic anisotropy (PMA) energy density ($K_s^{ISOC}$) of a given NM/FM interface is determined by the ISOC and interfacial orbital hybridization, i.e. $K_s^{ISOC}/t_{FM} \propto \xi (m_o^\perp - m_o^\parallel)$, where $t_{FM}$, $m_o^\perp$, $m_o^\parallel$ are the thickness, perpendicular orbital magnetic moment, and the in-plane orbital magnetic moment of the FM layer, and $\xi$ is the ISOC energy of the interface, respectively. It has also been well established that $m_o^\perp$ is localized at the first atomic layer of the FM adjacent to the interface [46,47], i.e. $m_o^\perp = m_{o,i}^\perp/t_{FM} + m_o^\parallel$, where $m_{o,i}^\perp$ is the $m_o^\perp$ value for the single FM interface layer and $m_o^\parallel$ is thickness-insensitive and reasonably approximates the bulk orbital magnetic moment value for the FM [47,48]. As a result, $K_s^{ISOC} \propto \xi m_{o,i}^\perp$ for the NM/FM interfaces. This approximately linear correlation between $K_s^{ISOC}$ and the ISOC strength has well explained recent experimental observations of two-magnon scattering [49,50], interfacial Dzyaloshinskii–Moriya interaction [51,52], and spin memory loss [39]. It has also been established that the Rashba constant ($\alpha_R$) follows $\alpha_R \propto \xi m_{o,i}^\perp/m_o^\parallel$ [53]. Provided that $m_{o,i}^\perp$ and $m_o^\parallel$ are approximately invariant with the layer thicknesses [47,48], $K_s^{ISOC}$ can be also a linear indicator of $\alpha_R$ of NM/FM interfaces.

To quantify $K_s^{ISOC}$ for the Ti/FeCoB interface, we first determined the total interfacial PMA energy density ($K_s$) of the two Co interfaces of the Ti/FeCoB/MgO samples using the relation $H_k \approx -4\pi M_s + 2K_s/M_s t_{FeCoB}$, where $M_s$ is the saturation magnetization and $H_k$ is the effective perpendicular magnetic anisotropy field of the FeCoB layer. From vibrating sample magnetometry measurements (see Fig. S1 in the Supplementary Materials [54]), the $M_s$ values of these SiO$_2$/Ti d/FeCoB 1 or 2 samples are determined to be 850-950 emu/cm$^3$. While this agrees with 870 emu/cm$^3$ for $(Bi_{1-x}Sb_x)_2Te_3$/Ti 2/FeCoB 1.4 [55], it is interesting to note that FeCoB has been reported to have a $M_s$ varying from 1260 emu/cm$^3$ on CoPt 24/Ti 0.8 [42] to 470 emu/cm$^3$ on Ta 5/Ti 5 [56]. The $H_k$ values for the PMA samples (Fig. 1(b)) are determined by fitting the in-plane field, $H_x$, dependence of the in-phase first HHVR ($V_{1\omega}$) to the sinusoidal electric field $E$ = 66.7 kV/cm applied to the Hall bar following the relation

$$V_{1\omega} = \pm V_{AH} \approx \pm V_{AH}(1-H_x^2/2H_k^2), \quad (1)$$

where $V_{AH}$ is the anomalous Hall voltage determined by measuring $V_{1\omega}$ as a function of $H_z$ (see Fig. 2(a)), $\theta \approx H/H_k <<$ 1 is the polar angle of the magnetic moment tilted away from the film normal direction, and the signs ± correspond to the initial magnetization states of $\pm M_z$, respectively. Here, the charge current flows in the $x$ direction, and z is the out-of-plane direction. We measured $H_x$ dependency of $V_{1\omega}$ for both the ±$M_z$ cases (see Fig. 2(b)) and averaged the two values of $H_k$ for the PMA Ti/FeCoB samples. For the in-plane samples, $dV_{1\omega}/dH_z$ is first determined under different in-plane bias fields $H_x$ (Fig. 2(c)), $H_k$ and $V_{AH}$ were then estimated by fitting the data to the relation

$$dV_{1\omega}/dH_z = V_{AH}/(-H_k+|H_x|). \quad (2)$$

As shown in Fig. 1(c), $K_s$ for the in-plane magnetized Ti/FeCoB samples first increases from ≈ 0.29 ± 0.03 erg/cm$^2$ at $d$ = 1.7 nm to 0.64 ± 0.02 erg/cm$^2$ at $d$ = 2.8 nm, and then gradually drops to 0.02 ± 0.02 erg/cm$^2$ at $d$ = 6.9 nm. First, the minimal $K_s$ value for the Ti 6.9/FeCoB 2/MgO should indicate a negligible ISOC at the top FeCoB/MgO interface and thus $K_s^{ISOC} \approx K_s$ for these Ti/FeCoB interfaces. This is an interesting observation as our previous Pt$_x$Pd$_{1-x}$/FeCoB 2-7/MgO [51] with high $M_s$ of 1200 emu/cm$^3$ show $K_s$ of 0.49 erg/cm$^2$ at the top FeCoB/MgO interfaces. Furthermore, the strong variation of $K_s$ and thus $K_s^{ISOC}$ suggests a strong tuning of the ISOC strength by the thickness of the Ti layer. The dependence of ISOC, as well as $M_s$, on the underlayers likely suggests a notable sensitivity of the magnetic order of the FeCoB to its underlayers. Note that some other HM/FM interfaces (e.g. Pt/Co [52]) also show significant change of $K_s$ at large HM thicknesses. *ab initio* calculations could be informative for understanding the microscopic mechanism of the interesting variation of ISOC and $M_s$ with $d$, it is, however, beyond the scope of this Letter. For the PMA Ti/FeCoB samples (red dots), $K_s^{ISOC}$ remains at a high value of 0.70± 0.01 erg/cm². As indicated by the blue lines in Fig. 1(c), $K_s^{ISOC}$ is ≈ 0.6 erg/cm² for a typical Pt/Co interface [39] and ≈ 0.24 erg/cm² for a typical Pt/FeCoB interface [50].

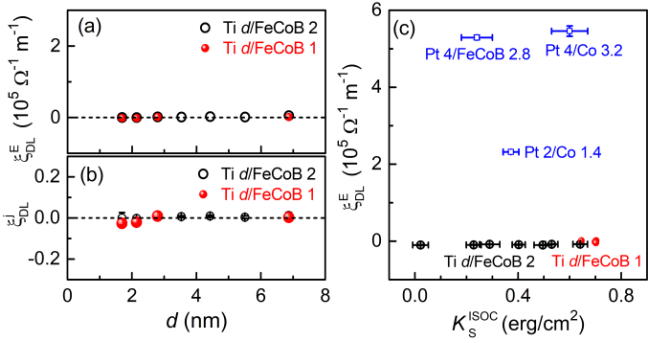

Fig. 3. (a) $\xi^E_{DL}$ vs $d$, (b) $\xi^j_{DL}$ vs $d$, and (c) $\xi^E_{DL}$ vs for Ti/FeCoB bilayers. The data for the Pt 4/FeCoB 2.8 [50], Pt 4/Co 3.2 [39], and Pt 2/Co 1.4 are plotted in (c) for comparison.

The dampinglike (fieldlike) SOT efficiencies per applied electric field, $\xi^E_{DL(FL)}$, and per unit bias current density, $\xi^j_{DL(FL)}$, of the Ti/FeCoB bilayers can be determined using HHVR measurements as [57]

$$\xi^E_{DL(FL)} = (2e/\hbar) \mu_0 M_s t_{FM} H_{DL(FL)}/E. \quad (3)$$
$$\xi^j_{DL(FL)} = (2e/\hbar) \mu_0 M_s t_{FM} H_{DL(FL)}/j. \quad (4)$$

where $H_{DL(FL)}$ is the dampinglike (fieldlike) effective SOT field, and $j$ the bias current density. For Ti 1.7-6.9/FeCoB 1 with PMA, $H_{DL(FL)} = -2\frac{\partial V_{2\omega}}{\partial H_{x(y)}}/\frac{\partial^2 V_{1\omega}}{\partial H^2_{x(y)}}$ [43], where the in-phase first HHVR ($V_{1\omega}$) and the out-of-phase second HHVR ($V_{2\omega}$) are parabolic and linear functions of in-plane magnetic fields $H_{x(y)}$ [Fig. 2(b)], respectively. Here, we do not apply the so-called "planar Hall correction" in analyzing the out-of-plane HHVR results for the reasons discussed in detail in the Supplementary Materials of Refs. [44] and [51]. For the in-plane magnetized Ti 1.7-6.9/FeCoB 2, $H_{DL(FL)}$ is determined from the angle-dependent in-plane HHVR measurements (Fig. 2(d)) and reaffirmed by the field-dependent in-plane HHVR measurements (Fig. S2 in the Supplementary Materials [54]). As shown in Fig. 2(d), $V_{2\omega}$ follows [45,51]

$$V_{2\omega} = (V_{DL} + V_{ANE}) \cos\varphi + V_{FL} \cos\varphi\cos2\varphi, \quad (5)$$

where $V_{DL} = -V_{AH}H_{DL}/2(H_{xy}+|H_k|)$ is the second HHVR to the dampinglike SOT, $V_{FL} = V_{PH}(H_{FL} + H_{Oe})/H_{xy}$ the second HHVR to the fieldlike SOT and Oersted field torque, $V_{ANE}$ the anomalous Nernst voltage, $H_{xy}$ the in-plane bias field, $\varphi$ the in-plane angle of $H_{xy}$ and thus the magnetization with respect to the current direction. $H_{DL}$ and $H_{FL}$ are determined from the slope of the linear fit of $V_{DL}$ vs $-V_{AH}/2(H_{xy}+|H_k|)$ [see Fig. 2(e)] and $V_{FL}$ vs $V_{PH}/H_{xy}$, respectively.

As summarized in Fig. 3(a), the obtained values of $\xi^E_{DL}$ of the Ti/FeCoB samples are negligibly small regardless of the Ti thickness, i.e. $|\xi^E_{DL}| < 0.04\times10^5$ Ω⁻¹ m⁻¹ for Ti/FeCoB. Although this value is slightly higher than the "intrinsic Berry curvature"-inuduced torque at $Ni_{81}Fe_{19}/CuO_x$ interface ($0.016\times10^5$ Ω⁻¹ m⁻¹)[34], it is still two orders of magnitude smaller than ≈ $5.5\times10^5$ Ω⁻¹ m⁻¹ for Pt 4/Co [50], Pt 4/FeCoB [39], and Pt 4/$Ni_{81}Fe_{19}$ bilayers [50] and thus does not represent an efficient charge-to-spin conversion at all. The corresponding vaues of $\xi^j_{DL} = \xi^E_{DL}\rho_{xx}$ are also very small ($|\xi^j_{DL}|< 0.025$) despite the high resistivity ($\rho_{xx}$) of the thin Ti layer ($\rho_{xx}$ decreases gradually from 1320 μΩ cm for $d = 1.7$ nm to 220 μΩ cm for $d = 6.9$ nm). We note that, previous Ti 1.2/NiFe/$Al_2O_3$ [58] and Ti 3 or 5/CoFeB/MgO samples [59], despite the uncharacterized ISOC strength, were found to have negligible dampinglike SOT, which agrees with our observation. The origin of this negligbly small dampinglike SOT in Ti/FeCoB bilayers is unclear. One of the possiblities is a very weak OHE in Ti. Note that recent theories and experiments have indicated that orbital current can be converted to spin current at an interface with strong SOC [23,24]. The bulk SHE of Ti is negligible and irrelevant to the small $\xi^E_{DL}$ of Ti/FeCoB because the $\xi^E_{DL} /\xi^E_{FL}$ ratio is < 0.5.

More importantly, the observed irrelevance of the torque efficiencies to the ISOC strength of the Ti/FeCoB interface (Fig. 3(c)) provide unambiguous evidence that there is no significant interfacial generation of spin current by Rashba-Edelstein effect and spin-orbit filtering in these samples. Note that the spin current generation and thus dampinglike SOT due to an interfacial Rashba-Edelstein or spin filtering effect, if any, should increase linearly with $\alpha_R$ [19,21] and thus $K_s^{ISOC}$. This result most likely indicates that the interfacial generation of spin current and SOTs by Rashba-Edelstein effect and spin filtering effect should be also absent in the Pt/Co and the Pt/FeCoB samples and other NM/FM heterostructures which have similar or lower ISOC strength as these Ti/FeCoB interfaces. Instead, the very strong dampinglike SOTs of ~$10^5$ Ω⁻¹ m⁻¹ in the Pt 4/FeCoB 2.8 [50], Pt 4/Co 3.2 [39], and Pt 2/Co 1.4 in Fig. 3(c) should be entirely attributed to the giant intrinsic SHE of the Pt [7,35,36].

The absence of any signficant ISOC generation of spin current appears to be a quite general phenomena rather than being just specific to the Ti/FeCoB system. First, this is consistent with the generalized theories that ISOC has negligible contribution to the dampinglike SOT via the 2D Rashba-Edelstein(-like) effect at metallic magnetic interfaces [25-28]. Experimentally, this is well supported by: (i) a previous spin Seebeck/ISHE study that spin current generation is absent at Bi/Ag/$Y_3Fe_5O_{12}$ and Bi/$Y_3Fe_5O_{12}$ interfaces prepared by differnet techniques [32]; (ii) the universal observation of strong scaling of the SOT efficiencies or the effective spin Hall angle with the resistivity and the layer thickness of HMs [7,36,37,51,57,60-62], topological insulators [8], nonmagnetic complex oxides [63];

and (iii) the clear dependence of inverse spin Hall voltage on the spin-mixing conductance of the interfaces of magnetic oxides (e.g. $Y_3Fe_5O_{12}$) in spin Seebeck [64], spin pumping [65], and spin magnetoresistance processes [66]. Very importantly, this finding reaffirms that the rapid linear reduction of dampinglike SOT with increaseing ISOC in annealed Pt/Co and $Au_{1-x}Pt_x$/Co [39] is because the effective consequence of ISOC is not a strong negative interfacial dampinglike torque but rather a substantial loss of spin Hall spin current to the lattice (i.e. spin memory loss).

Finally, we find that the fieldlike torque at the Ti/FeCoB interfaces is sizable but still small, $\xi_{FL}^E \approx -0.1 \times 10^5$ $\Omega^{-1}$ $m^{-1}$ (see Fig. S3 in the supplementary Materials [54]), and shows no obvious correlation to $K_s^{ISOC}$ and thus $\alpha_R$. This is in line with our consistent observation of lack of any distinguishable positive correlation between $\xi_{FL}^E$ and $K_s^{ISOC}$ in Pt-based spin Hall metal/FM interfaces [51]. We note that a weak fieldlike torque was also observed previously in Ti 1.2/NiFe/$Al_2O_3$ [58], however, that work did not verify the mechanism of that fieldlike torque by any quantitative correlation to the strength of the ISOC.

In summary, we have presented unambiguous evidence for the absence of any significant spin current generation at the Ti/FeCoB interfaces. By varying the layer thicknesses, the strength of the ISOC at Ti/FeCoB interface is tuned from being negligibly small to being greater than that of typical Pt/FeCoB and Pt/Co interfaces. From this ability, we find that both dampinglike and fieldlike SOTs in the Ti/FeCoB bilayers remain negligibly small compared to that from the bulk SHE of Pt-based NM/FM systems (more than two orders of magnitude smaller), regardless of the Ti thickness, the ISOC strength, and the type of magnetic anisotropy. Our results, in combination with the existing theories [25-28] and experiments [7,8,36,37,51,57,60-66], most likely suggest the absence of any significant spin current generation by spin-orbit filtering effect and the Rashba-Edelstein-like effect at other magnetic interfaces. This result also indicates negligible bulk SHE and OHE in Ti. Our results provide important information for the fundamental understanding of spin current generation, charge-spin conversion, and SOTs at magnetic interfaces.

This work was supported in part by the Office of Naval Research (N00014-15-1-2449), in part by the NSF MRSEC program (DMR-1719875) through the Cornell Center for Materials Research, and in part by the Defense Advanced Research Projects Agency (USDI D18AC00009). The devices were fabricated at the Cornell NanoScale Facility, an NNCI member supported by NSF Grant ECCS-1542081.

**Supplementary Materials for**

**Absence of significant spin current generation in Ti/FeCoB bilayers with strong interfacial spin-orbit coupling**

Lijun Zhu[*], Robert A. Buhrman

*Cornell University, Ithaca, New York 14850, USA*

*lz442@cornell.edu

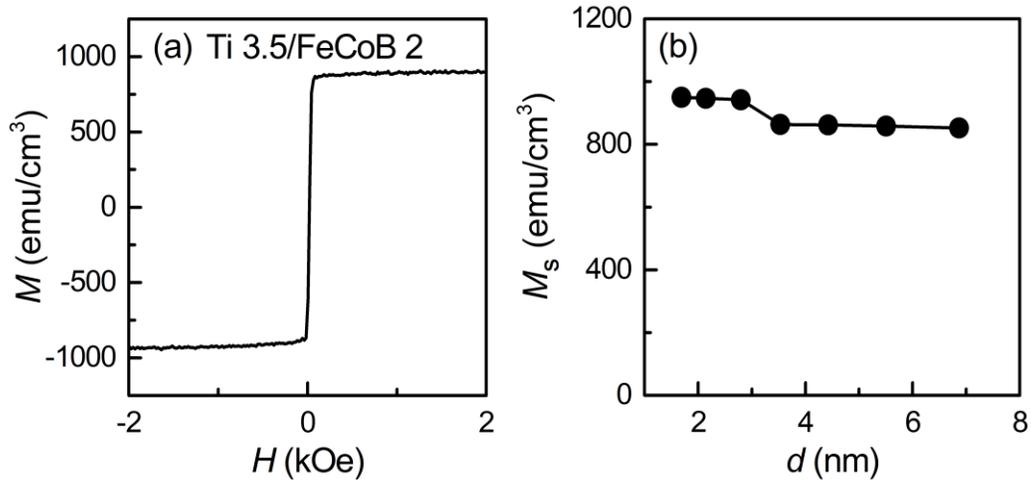

Fig. S1 (a) In-plane magnetization vs in-plane magnetic field for Ti 3.5 nm/FeCoB 2. (b) Saturation magnetization of Ti $d$/FeCoB 2 samples plotted as a function of the Ti layer thickness.

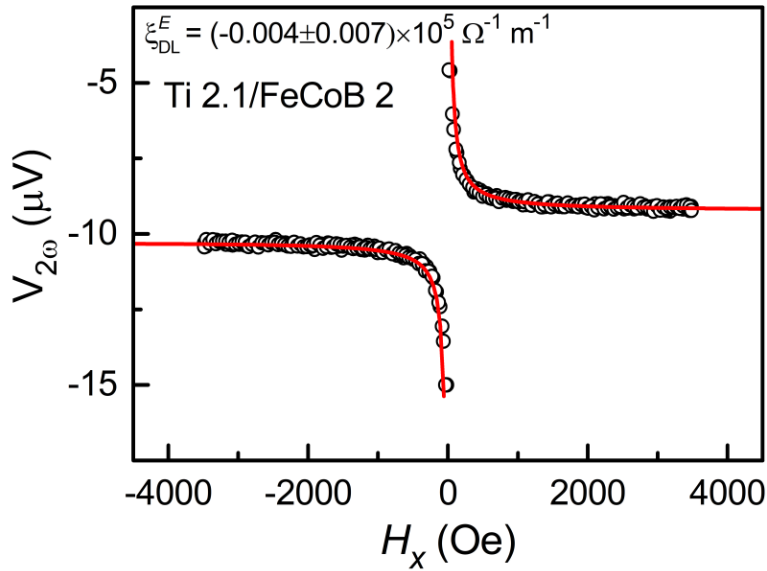

Fig. S2 In-plane magnetic field dependent harmonic Hall voltage response measurement. The second harmonic Hall voltage of Ti 2.1/FeCoB 2 plotted as a function of in-plane magnetic field along current direction ($H_x$). The solid lines represent the best fits of the data to Eq. [5], from which we obtain a negligible dampinglike spin-orbit torque efficiency per electric field, in consistence with the angle-dependent HHVR results in Fig. 2 of the main text.

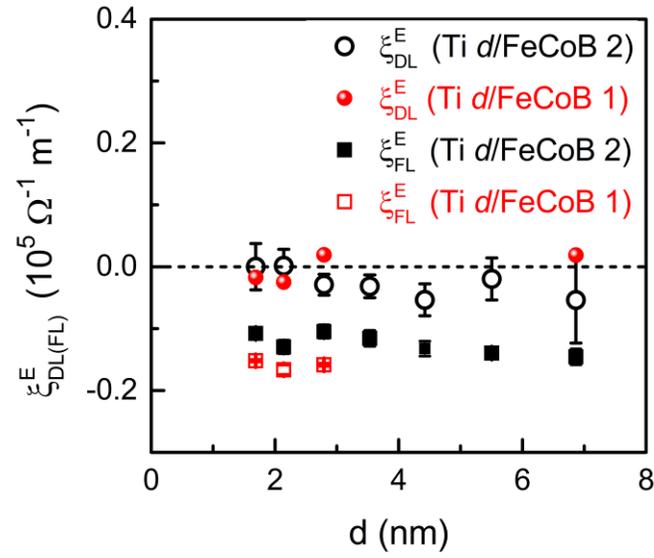

Fig. S3. Damplinglike and fieldlike spin-orbit torques per applied electric field for the perpendicularly magnetized bilayers of Ti $d$/FeCoB 1 and in-plane magnetized bilayers of Ti $d$/FeCoB 2.